\documentclass[12pt]{article}

\usepackage[dvips]{graphicx}
\usepackage{amssymb}
\usepackage{amsmath}
\usepackage{latexsym}

\def\Slash#1{{\ooalign{\hfil$#1$\hfil\crcr\hfil$/$\hfil}}}

\begin{document}

\begin{titlepage}
\begin{center}
{\small
\hfill TU-870 \\[-1mm]
\hfill KEK Preprint 2010-21 \\[-1mm]
\hfill OU-HET 672/2010 \\[-1mm]
\hfill UT-HET 037 \\[-1mm]
\hfill KUNS-2276 \\
}

\vspace{1.0cm}
{\large\bf Extra dimensions and Seesaw Neutrinos \\[1mm]
at the International Linear Collider}
\vspace{1.2cm}

{\bf Tomoyuki Saito}$^{(a)}$,
{\bf Masaki Asano}$^{(a)}$,
{\bf Keisuke Fujii}$^{(b)}$,
{\bf Naoyuki Haba}$^{(c)}$, \\
{\bf Shigeki Matsumoto}$^{(d)}$,
{\bf Takehiro Nabeshima}$^{(d)}$,
{\bf Yosuke Takubo}$^{(a)}$, \\
{\bf Hitoshi Yamamoto}$^{(a)}$,
and
{\bf Koichi Yoshioka}$^{(e)}$

\vspace{1cm}

{\it%
$^{(a)}${Department of Physics, Tohoku University, Sendai 980-8578, Japan} \\
$^{(b)}${High Energy Accelerator Research Organization (KEK), Tsukuba 305-0801, Japan} \\
$^{(c)}${Department of Physics, Osaka University, Toyonaka 560-0801, Japan} \\
$^{(d)}${Department of Physics, University of Toyama, Toyama 930-8555, Japan} \\
$^{(e)}${Department of Physics, Kyoto University, Kyoto 606-8502, Japan}
}

\vspace{1.5cm}

\abstract{We study the capability of the international linear collider (ILC) to probe extra dimensions via the seesaw mechanism. In the scenario we study, heavy Kaluza-Klein neutrinos generate tiny neutrino masses and, at the same time, have sizable couplings to the standard-model particles. Consequently, a Kaluza-Klein tower of heavy neutrinos ($N$) can be produced and studied at the ILC through the process: $e^{+} e^{-} \to \nu N$ followed by $N \to W \ell$ decay. We show that the single lepton plus two-jets final states with large missing energy from this signal process will provide a good opportunity to measure the masses and cross sections of Kaluza-Klein neutrinos up to the third level. Furthermore, the neutrino oscillation parameters can be extracted from the flavor dependence of the lowest-mode signals, which give us information about the origin of low-energy neutrino masses.}

\end{center}
\end{titlepage}

\setcounter{footnote}{0}


\section{Introduction} 
\label{sec:intro}

The detailed structure of lepton sector has been gradually revealed by the recent neutrino oscillation experiments~\cite{neu_review,neu_analysis}. The smallness of neutrino masses is one of the most important clues to find new physics beyond the standard model (SM). The seesaw mechanism naturally leads to small neutrino masses through heavy particles being coupled with ordinary neutrinos. In Type I seesaw scheme~\cite{seesaw}, the introduction of right-handed neutrinos implies intermediate mass scales to have light neutrino masses of order eV, and hence these heavy states are almost decoupled in low-energy effective theory. Alternatively, TeV-scale right-handed neutrinos are also viable, which in turn means much smaller couplings to the SM sector and their signals would not be captured in future collider experiments such as the Large Hadron Collider (LHC). It is therefore difficult to simultaneously realize tiny neutrino masses and detectably large interactions among right-handed neutrinos and the SM particles within the framework of four-dimensional Type I seesaw.

In the previous work by a part of the authors~\cite{HMY}, it was pointed out that the difficulty is overcome by a simple extension of the SM\@. We have considered a five-dimensional theory where all the SM fields are confined in a four-dimensional space-time while right-handed neutrinos propagate in the bulk of extra-dimensional space~\cite{DDG, nuR_ExD}. With bulk Majorana mass, TeV-scale right-handed neutrinos can generate a tiny scale of neutrino masses through the seesaw mechanism and simultaneously have sizable couplings to the SM leptons and gauge bosons. The previous work focused on the tri-lepton signal with large missing transverse energy $pp \to \ell^\pm\ell^\mp\ell^\pm\nu (\bar\nu)$. This process is expected to be detectable at the LHC because only a small fraction of SM processes contributes to the background against the signals. It was shown that the observation of right-handed neutrinos is possible, though it is limited only to the lightest Kaluza-Klein (KK) mode of right-handed neutrinos and the mechanism itself seems difficult to be confirmed at the LHC. 

In this paper, we will investigate how this scenario can be observed at the International Linear Collider (ILC). The ILC is the future electron-positron linear collider for the next generation of high-energy frontier physics. At the ILC, electrons and positrons are accelerated by two opposing linear accelerators installed in an about 30 km long underground tunnel, and are brought into collision with a center of mass energy of $500\;\text{GeV}$-$1\;\text{TeV}$\@. The clean experimental environment of the ILC due to the electron-positron collider gives us an opportunity to obtain more information about the property of right-handed neutrinos and their interactions. We focus on the two-jet and single lepton signal with large missing  energy, $e^+ e^- \to \nu N$ ($N \to \ell W,\, W \to q\bar{q}$), which can be efficiently reconstructed at the ILC and allows us to reveal not only the lightest but also higher KK-excited states of right-handed neutrinos. As a result, the ILC experiment can completely confirm that the scenario is based on the extra-dimensional theory. Furthermore, by observing the flavor dependence of interactions between right-handed neutrinos and SM particles, it is also possible to confirm the mechanism to generate the neutrino masses. We performed the analysis of the signal in the various mass hierarchies of neutrino masses (the normal, inverted, and degenerate cases). 

This article is organized as follows. In the next section, we briefly review the five-dimensional theory with right-handed neutrinos. The simulation framework such as a representative point in the parameter space of the model and the simulation tools used in our study are presented in Section~\ref{sec:setup}. The details of analysis to observe the right-handed neutrinos at the ILC are discussed in Section~\ref{sec:analysis}, where we show expected measurement accuracies of the masses of right-handed neutrinos and their production cross-sections at the center of mass energy  of 500~GeV and 1~TeV\@. In Section~\ref{sec:discussion}, we will discuss how the mechanism of neutrino-mass generation can be confirmed at the ILC. Section~\ref{sec:summary} is devoted to summary.

\section{Physics Model}
\label{sec:Model}

We consider the five-dimensional theory on $S^1/Z_2$ with the $S^1$-radius $R$. The coordinates are denoted by $(x^\mu, y)$ where the $y$ direction is compactified to the line segment $y = [0, \pi R]$. The standard-model fields are confined on the four-dimensional boundary at $y=0$. In addition, we introduce the bulk gauge-singlet fermions ${\cal N}_i$ ($i=1,2,3$) and assign the even $Z_2$ parity to their upper components, i.e.\ ${\cal N}_i(-y) = \gamma_5\,{\cal N}_i(y)$, so that they contain three-generation right-handed neutrinos as zero modes.\footnote{We do not consider the possibility of generation-dependent parity assignment~\cite{HWY}.} The kinetic and mass terms are given by
\begin{eqnarray}
  {\cal L}
  \;=\;
  i\overline{{\cal N}_i} \Slash{D} {\cal N}_i
  -
  \frac{1}{2} \Big[\,\overline{{\cal N}^c_i}
  (M\gamma_5 + M')_{ij}
  {\cal N}_j + \text{h.c.}\Big].
\end{eqnarray}
The conjugated spinor is defined as ${\cal N}^c = \gamma_3 \gamma_1 \overline{\cal N}^{\rm t}$. It is easy to also write down the Dirac mass $m_d$ if one introduces a $Z_2$-odd function which could originate from some field expectation value. In this paper, we take $m_d = M' = 0$, first considered in Ref.~\cite{DDG} as a simple example of higher-dimensional seesaw models. We also have the mass term between bulk and boundary fields:
\begin{eqnarray}
  {\cal L}_m
   \;=\;
  \overline{{\cal N}_i} P_L m_{ij} L_j \,
  \delta(y) + {\rm h.c.},
  \label{boundary}
\end{eqnarray}
generated after the electroweak symmetry breaking. The boundary fermions $L_i$ contain the left-handed neutrinos $\nu_i$. Hereafter we take the basis where $M$ is generation diagonalized, and will drop other generation indices for notational simplicity.

The bulk fermions are KK expanded by Majorana fermions $\Psi_{R, L}^n$ as
\begin{eqnarray}
  {\cal N}(x, y)
  \;=\;
  \sum \limits_{n=0} \chi^n_R(y) P_R \Psi_R^n(x)
  +
  \sum \limits_{n=1} \chi^n_L(y) P_L \Psi_L^n(x).
\end{eqnarray}
The wavefunctions $\chi_{R, L}^n$ are normalized so that the KK-mode kinetic terms are canonical in four dimensions. The low-energy neutrinos (e.g.\ singlet fermions) come from the boundary neutrinos and the KK modes: $(\nu, \Psi_R^0, \Psi_R^1, \Psi_L^1, \Psi_R^2, \Psi_L^2, \cdots) \equiv (\nu, N)$. By integrating over the fifth dimension, we obtain the Majorana mass matrix in four-dimensional effective theory, explicitly given by $\frac{1}{2} (\overline{\nu^c} \;\, \overline{N^c}) P_L {\cal M} \big( \begin{smallmatrix} \nu\\[.5mm] N\end{smallmatrix} \big) + \text{h.c.}$,
\begin{eqnarray}
\qquad\qquad
  {\cal M} \;=\; \left(%
  \begin{array}{c|cccc}
    0 & \,m_0^{\rm t} & \,m_1^{\rm t} & 0 & \cdots \\
    \hline
    m_0 & M_{R_{00}}^* & M_{R_{01}}^* & M_{K_{01}}^{} & \cdots \\[1mm]
    m_1 & M_{R_{10}}^* & M_{R_{11}}^* & M_{K_{11}}^{} & \cdots \\[1mm]
    0 & M_{K_{10}}^{\rm t} & M_{K_{11}}^{\rm t} & M_{L_{11}}^{} & 
    \cdots \\[1mm]
    \vdots & \vdots & \vdots & \vdots & \ddots
  \end{array}\right)
  \;\equiv\; \left(
  \begin{array}{c|ccc}
    & & M_D^{\rm t} & \\
    \hline
    & & & \\
    \!\!M_D & ~ & M_N  & \\
    & & &
  \end{array}\right),
\end{eqnarray}
where the boundary, KK-, and Majorana 
masses ($m_n$, $M_K$, and $M_{R,L}$) are
\begin{alignat}{2}
  m_n \;&=\, \chi^n_R(0)m\,,& \qquad
  M_{R_{mn}} &= \int_{-\pi R}^{\pi R}\!\!dy\,
  \chi^m_R M \chi^n_R\,, \nonumber \\
  M_{K_{mn}} &= \int_{-\pi R}^{\pi R}\!\!dy\,
  \chi^m_R\partial_y\chi^n_L\,,& \qquad
  M_{L_{mn}} &= \int_{-\pi R}^{\pi R}\!\!dy\,
  \chi^m_L M \chi^n_L\,.
\end{alignat}
It is noted that $M_{K_{mn}}$ becomes proportional to $\delta_{mn}$ when $\chi_{R,L}^n$ are the eigenfunctions of bulk equations of motion, and $M_{R_{mn}}$, $M_{L_{mn}}$ are also proportional to $\delta_{mn}$ for a constant ($y$-independent) mass parameter $M$ due to the normalization conditions.

\subsection{Seesaw and Electroweak Lagrangian}

We further implement the seesaw operation assuming ${\cal O}(m_n) \ll {\cal O} (M_{R, L, K})$ and find the induced Majorana mass matrix for three-generations light neutrinos 
\begin{eqnarray}
  M_\nu \;=\; -M_D^{\text{t}}M_N^{-1}M_D^{}.
  \label{seesaw_mass}
\end{eqnarray}
It is useful for later discussions to write down the electroweak Lagrangian in the basis where all the mass matrices are generation-diagonalized. The neutrino interactions to electroweak gauge bosons are given in terms of these mass eigenstates $(\nu_d,N_d)$: 
\begin{eqnarray}
  {\cal L}_g &=& \frac{g}{\sqrt{2}}\Big[W_\mu^\dagger\,
  \bar e\gamma^\mu U_{\rm MNS} P_L \big(\nu_d+VN_d\big)+\text{h.c.}\Big]
  \nonumber \\
  && \qquad\qquad 
  +\frac{g}{2\cos\theta_W}Z_\mu\big(\bar\nu_d+
  \bar N_d V^\dagger\big)\gamma^\mu P_L\big(\nu_d+VN_d\big),\;
  \label{Lgauge}
\end{eqnarray}
where $W$ and $Z$ are the electroweak gauge bosons and $g$ is the $SU(2)_{\rm weak}$ gauge coupling constant. The spinors $\nu_d$ are three light neutrinos for which the seesaw-induced mass matrix~\eqref{seesaw_mass} is diagonalized 
\begin{eqnarray}
  M_\nu \;=\; U_\nu^*\,M_\nu^d\,U_\nu^\dagger,  \qquad
  U_\nu\,\nu_d \;=\; \nu-M_D^\dagger M_N^{-1\,*}N,
  \label{nud}
\end{eqnarray}
and $N_d$ denote the infinite numbers of neutrino KK modes for which the bulk mass matrix $M_N$ is diagonalized both in the generation and KK-mode spaces by a unitary matrix $U_N\,$:
\begin{eqnarray}
  M_N \,=\; U_N^*\,M_N^d\,U_N^\dagger,  \qquad
  U_N N_d\ \,=\; N+M_N^{-1}M_D^{}\,\nu.
  \label{Nd}
\end{eqnarray}

The lepton mixing matrix measured in neutrino oscillation experiments is given by $U_{\rm MNS}=U_e^\dagger U_\nu$ where $U_e$ is the left-handed rotation matrix for diagonalizing the charged-lepton Dirac masses. It is interesting to find in~\eqref{Lgauge} that the model-dependent parts of electroweak gauge vertices are governed by a single matrix $V$ defined as
\begin{eqnarray}
  V \;=\; U_\nu^\dagger M_D^\dagger M_N^{-1\,*}U_N.
\end{eqnarray}
When one works in the basis where the charged-lepton sector is flavor diagonal, $U_\nu$ is fixed by the neutrino oscillation matrix.

The neutrinos also have the Yukawa couplings to the Higgs doublet $H$ in the four-dimensional boundary, from which the Dirac mass~\eqref{boundary} is generated; 
\begin{eqnarray}
  {\cal L}_h \;=\; f\tilde H^\dagger\overline{\cal N}P_LL\,\delta(y)
  +\text{h.c.},
\end{eqnarray}
where $\tilde H = \epsilon H^*$. The doublet Higgs $H$ has a non-vanishing expectation value $v$ and its fluctuation $h(x)$ in the lower component. After integrating out the fifth dimension and diagonalizing mass matrices, we have the Yukawa interaction
\begin{eqnarray}
  {\cal L}_h \;=\; \frac{h}{v}\sum_n
  \overline{\Psi_R^n}\, m_nP_LU_\nu(\nu_d+VN_d) +\text{h.c.},
\end{eqnarray}
and $\Psi_R^n$ are determined by the mass eigenstates through 
Eqs.~\eqref{nud} and \eqref{Nd}.

\subsection{Observable Seesaw}

The interactions between heavy neutrinos and SM fields are described by the mixing matrix $V$ both in the gauge and Higgs vertices. The $3\times\infty$ matrix $V$ is determined by the mass parameters of neutrinos in the original Lagrangian ${\cal L} + {\cal L}_m$. The matrix elements in $V$ have the experimental upper bounds from electroweak physics. Another important constraint on $V$ comes from the low-energy neutrino experiments, namely, the seesaw-induced mass should be on the order of eV scale, which in turn specifies the scale of heavy neutrino mass $M_N$. This can be seen from the definition of $V$ by rewriting it with the light and heavy neutrino mass eigenvalues 
\begin{eqnarray}
  V \;=\; i(M_\nu^d)^{\frac{1}{2}}X(M_N^d)^{-\frac{1}{2}},
  \label{V}
\end{eqnarray}
where $X$ is an arbitrary $3\times\infty$ matrix with $X X^{\rm t} = 1$. Therefore one naively expects that, with a fixed order of $M_\nu^d \sim 10^{-1}\,\text{eV}$ and $|V| \gtrsim 10^{-2}$ for the discovery of experimental signature of heavy neutrinos, their masses should be very light and satisfy $M_N^d\lesssim$~keV (this does not necessarily mean that the seesaw operation is not justified as $M_\nu^d$ is fixed). The previous collider studies of TeV-scale right-handed neutrinos~\cite{TeVRH} did not satisfy the seesaw relation~\eqref{V} and have to rely on some assumption for suppressing the necessarily-induced (large) mass $M_\nu$; for example, the neutrino mass matrix must have a singular generation structure, otherwise it leads to the decoupling of heavy neutrinos from collider physics.

A possible scenario for observable heavy neutrinos is to take a specific value of bulk Majorana mass~\cite{HMY} so that the lepton number is recovered in low-energy effective theory. In this paper we assume that bulk Dirac mass vanishes, but it is easy to include it by attaching wavefunction factors in the following formulas. The equations of motion without bulk Dirac mass are solved by simple one-dimensional oscillators, and the neutrino mass matrices are found
\begin{alignat}{2}
  m_n \,\;&=\, \frac{m}{\sqrt{2^{\delta_{n0}}\pi R}}\,,& \qquad\;\;
  M_{R_{mn}} &=\; M\delta_{mn}, \nonumber \\[.5mm]
  M_{K_{mn}} &=\; \frac{n}{R}\delta_{mn}\,,& \qquad\;\;
  M_{L_{mn}} &=\; M\delta_{mn}.
\end{alignat}
From these matrices, we find the seesaw-induced mass matrix $M_\nu$ and the light neutrino mixing with heavy modes as follows:
\begin{eqnarray}
  M_\nu &=& m^{\rm t}\frac{1}{2\tan(\pi R|M|)}m,  \\[1mm]
  \nu &=& U_\nu\nu_d +\frac{1}{\sqrt{2\pi R}}\,m^\dagger
  \Bigg[ \sum_{n=0}\frac{1}{|M|+\frac{n}{R}}N_d^{2n+1}
  +\sum_{n=1}\frac{i}{|M|-\frac{n}{R}}N_d^{2n} \Bigg].
  \label{LRmixing}
\end{eqnarray}
The heavy mass eigenstate $N_d^{2n}$ ($N_d^{2n+1}$) are the Majorana fermions with masses $\frac{n}{R} - |M|$ ($\frac{n}{R} + |M|$). In the seesaw formula, the effect of infinitely many numbers of KK neutrinos appears as the factor $\pi R|M|/\tan(\pi R|M|)$. An interesting case is that $|M|$ takes a specific value $|M|\simeq\alpha/R$ where $\alpha$ is some half integer~\cite{DDG}: the seesaw-induced mass $M_\nu$ becomes tiny as a result of KK-mode summation (not only suppressed by the Majorana mass scale). On the other hand, the heavy-mode interaction is not suppressed unlike the above naive speculation. These facts realize the situation that right-handed neutrinos in the seesaw mechanism are observable at sizable rates in future collider experiments such as the LHC and ILC.

In the KK-mode picture, the mass spectrum is vector-like with a half integer $\alpha$ and no chiral zero mode exists; for $\alpha=1/2$, the mass eigenstates $N_d$ compose the Dirac fermions $N_n$ ($n=1,2,\cdots$) with masses $M_n$,
\begin{eqnarray}
  N_n \,=\, \frac{1}{\sqrt{2}}(N_d^{2n}-iN_d^{2n-1}), \qquad\quad
  M_n\,=\,\frac{2n-1}{2R}.
\end{eqnarray}
As a result, the lepton number is preserved in the KK-mode sector and their contributions to the seesaw-induced mass vanish. The model given above is an illustrative example for accessible seesaw neutrinos. While there are many other possibilities for extra-dimensional seesaw, they are supposed to have a common mass matrix structure as a key ingredient for tiny neutrino masses and observability, which could be seen from the operator analysis in low-energy effective theory. It would therefore be reasonable that the above model is used as a representative of extra-dimensional neutrinos. With the use of Eq.~\eqref{LRmixing}, the weak interaction of KK Dirac fermions, which is relevant to the collider study, turns out to be 
\begin{eqnarray}
 {\cal L}_{\rm int} 
 &=&
 -\frac{g}{\sqrt{2}} \sum_{n=1} \frac{1}{\pi R M_n}
 W_\mu^\dagger\,\bar{e}\gamma^\mu U_{\rm MNS} 
 \bigg(\frac{2M_\nu^d}{\delta_M}\bigg)^\frac{1}{2}\! P_L N_n
 \nonumber \\
 && \quad
 -\frac{g}{2\cos\theta_W} \sum_{n=1} \frac{1}{\pi R M_n}
 Z_\mu\,\bar\nu_d\gamma^\mu 
 \bigg(\frac{2M_\nu^d}{\delta_M}\bigg)^\frac{1}{2}\! P_L N_n
 \nonumber \\
 && \qquad
 -\sum_{n=1} \frac{1}{\pi R v} h \bar\nu_d 
 \bigg(\frac{2M_\nu^d}{\delta_M}\bigg)^\frac{1}{2}\! P_R N_n 
 \,+\text{h.c.}\,.
 \label{KKinteraction}
\end{eqnarray}
Here $\delta_M\equiv\frac{1}{2R}-|M|$ characterizes the scale of small
neutrino masses.

Similarly to the seesaw neutrino mass, heavy neutrinos do not give sizable contributions to lepton-number-violating processes such as the like-sign di-leptons in the final states~\cite{dileptons}. In the previous work, we have analyzed the LHC signature of the above model focusing on the (lepton-number-conserving) tri-lepton signal with large missing transverse energy~\cite{HMY} (see also~\cite{BMOZ}). It was found that the model gives enough excessive tri-lepton events beyond the SM background in a wide region of parameter space, and the LHC would discover the signs of neutrino mass generation and extra dimensions, while the analysis only included the contribution from the 1st KK-excited mode. In the following sections, we will perform the ILC study of the same setup, in particular, the observation of higher KK neutrino modes and their interactions to the SM particles.

\section{Representative Points and Simulation Tools}
\label{sec:setup}

\subsection{Constraints on Yukawa Couplings}

\begin{table}[t]
\begin{center}
\begin{tabular}{c|ccc} \hline 
Hierarchy & $m_{\nu 1}$ & $m_{\nu 2}$ & $m_{\nu 3}$ \\ \hline
(N) & 0 & $\Delta m_{21}$ & $\Delta m_{21} + \Delta m_{32}$ \\ 
(I) & $\Delta m_{32} - \Delta m_{21}$ & $\Delta m_{32}$ & 0 \\ 
(D) & $m_{\rm tot}$ & ~ $m_{\rm tot} + \Delta m_{21}$ 
& ~ $m_{\rm tot} + \Delta m_{21} + \Delta m_{32}$ \\ \hline
\end{tabular}
\caption{\small Three types of neutrino mass hierarchies used in our simulation study: (N), (I), and (D) correspond to the normal, inverted, and degenerate mass spectrum, respectively.} 
\label{table:hierarchy}
\end{center}
\end{table}   

Before going to discuss representative points used in our simulation study, we summarize the neutrino mass and mixing matrices which are mandatory to determine the flavor structure of Yukawa interaction. The two matrices are parameterized as
\begin{eqnarray}
 M_\nu^d &=&
 \left(\begin{array}{ccc}
   m_{\nu_1} & & \\
   & m_{\nu_2} & \\
   & & m_{\nu_3} 
 \end{array}\right),
 \qquad\quad
 \phi \;=\;
 \left(\begin{array}{ccc}
   e^{i\varphi_1} & & \\
   & e^{i\varphi_2} & \\
   & & 1
  \end{array}\right),
 \nonumber \\
 U_{\rm MNS} &=&
 \left(\begin{array}{ccc}
   c_{12} c_{13} &
   s_{12} c_{13} &
   s_{13} e^{-i\delta} \\
   -s_{12} c_{23} - c_{12} s_{23} s_{13} e^{i\delta} &
   c_{12} c_{23} - s_{12} s_{23} s_{13} e^{i\delta} &
   s_{23} c_{13} \\
   s_{12} s_{23} - c_{12} c_{23} s_{13} e^{i\delta} &
   - c_{12} s_{23} - s_{12} c_{23} s_{13} e^{i\delta} &
   c_{23} c_{13}
 \end{array}\right)\phi\,,
\end{eqnarray}
where $s_x$ $(c_x)$ means $\sin \theta_x$ ($\cos \theta_x$). The Dirac and Majorana phases are denoted by $\delta$ and $\varphi_{1,2}$, respectively. The neutrino mass differences and the generation mixing parameters have been measured at the neutrino oscillation experiments~\cite{neu_analysis}. We take their typical values; $\Delta m_{21} \equiv m_{\nu_2} - m_{\nu_1} \simeq 9 \times 10^{-3}$ eV, $\Delta m_{32} \equiv |m_{\nu_3} - m_{\nu_2}| \simeq 5 \times 10^{-2}$ eV, $s_{12} \simeq 0.56$, $s_{23} \simeq 0.71$, and $s_{13} \leq 0.22$. The mass spectrum is allowed to have three types of hierarchies shown in Table~\ref{table:hierarchy}, where we define $m_{\rm tot} = (0.67~\text{eV} -2\Delta m_{21} - \Delta m_{32})/3$, considering the cosmological bound $\sum_i m_{\nu_i} \leq 0.67$ eV~\cite{mtotal}.

Since the scenario we are studying also affects several physical observables such as the flavor-changing processes of charged leptons, it is important to take account of constraints on the neutrino Yukawa coupling to have proper representative points. By integrating out all the heavy KK fermions from the Lagrangian~\eqref{KKinteraction}, we obtain the following dimension 6 operator ${\cal O}^{(6)}$, which contributes to the flavor-changing neutral current;
\begin{eqnarray}
  {\cal O}^{(6)} \,=\, \frac{\pi^2 R^2}{2}
  \big(\bar{L} \tilde H\big) f^\dagger f \Slash{\partial}
  \big(\tilde H^\dagger L\big), \qquad\;\;
  f \,=\, \frac{2}{\pi Rv}\delta_M^{\,-\frac{1}{2}}\,YU^\dagger_{\rm MNS},
\end{eqnarray}
where $Y$ is the $3\times3$ orthogonal matrix which generally comes in reconstructing high-energy quantities from the observable ones~\cite{CI}. That corresponds to the matrix $X$ in~\eqref{V}. The coefficient of the operator receives phenomenological constraints as shown in Ref.~\cite{lowene}, and then each component of the Yukawa couplings is restricted by comparing theoretical predictions with experimental data.

\subsection{Representative Points}

We choose the representative points with $M_i=M\times {\bf 1}$, namely the right-handed neutrino masses are degenerate in the flavor space, and also assume that $Y$ is a real orthogonal matrix. As a result, the operator ${\cal O}^{(6)}$ is found to be
\begin{eqnarray}
  {\cal O}^{(6)} \,=\,
  \frac{2}{v^2\delta_M} \big(\bar{L}\tilde H\big)
  U_{\rm MNS} M_\nu^d U_{\rm MNS}^\dagger \Slash{\partial}
  \big(\tilde H^\dagger L\big).
\end{eqnarray}
To satisfy experimental constraints from this operator and become small e-$\mu$ component, we take the lepton mixing matrix $U_{\rm MNS}$ as shown in Table~\ref{table:MNS} for each case of neutrino mass hierarchy. 

\begin{table}[t]
\begin{center}
\begin{tabular}{c|cccc} \hline 
Hierarchy & $s_{13}$ & $\delta$ & $\varphi_1$ & $\varphi_2$ \\ \hline
(N)       & 0.07     & $\pi$    & 0           & 0           \\
(I)       & 0.09     & 0        & 0           & 0           \\
(D)       & 0.04     & $\pi$    & 0           & 0           \\ \hline
\end{tabular}
\caption{\small The representative points for $U_{\rm MNS}$.(N), (I), and (D) correspond to the normal, inverted, and degenerate mass spectrum, respectively.}
\label{table:MNS}
\end{center}
\end{table}   

Interestingly, the coefficient of ${\cal O}^{(6)}$ depends only on the parameter $\delta_M$, which turns out to be constrained as follows in each pattern of neutrino mass hierarchy: 
\begin{alignat}{2}
  \delta_M &\;\geq\; 3.3~{\rm eV} & \qquad & \text{for (N)}, \\
  \delta_M &\;\geq\; 4.4~{\rm eV} & \qquad & \text{for (I)}, \\
  \delta_M &\;\geq\;\; 24~{\rm eV} & \qquad & \text{for (D)}.
\end{alignat}
We will set $\delta_M$ to the most optimistic value, namely these lower bounds, in the following simulation study. The new study~\cite{Antusch:2008tz} may be given the stronger constraints.The change of the constraints can be put in the change of $\delta_M$ bound.It is straightforward to extend them to larger values because collider signals such as the production cross sections of heavy KK neutrinos are simply proportional to $1/\delta_M$, though the discovery of the signal will be difficult.

The compactification radius $R$ is not relevant to the constraint from the operator ${\cal O}^{(6)}$. It is however limited by the LEP experiment since the Dirac masses of KK neutrinos are given by $M_n=(2n-1)/2R$. It is easy to confirm that the constraint is not so severe if $1/R > 200$ GeV, and we thus use the value
\begin{eqnarray}
  1/R \,=\, 300~{\rm GeV},
\end{eqnarray}
as a representative point of $1/R$ in our simulation study.

\subsection{Simulation Tools}

\begin{table}
 \center{
  \begin{tabular}{lcr}
   \hline
   Detector & Performance & Coverage \\
   \hline
   Vertex detector &
   $\delta_{b} \leq 5 \oplus 10/ p \beta \sin^{3/2}\theta$ ($\mu$m) &
   $|\cos\theta| \leq 0.93$
   \\
   Central drift chamber &
   $\delta p_{t}/p_{t}^{2} \leq 5 \times 10^{-5}$ (GeV/c)$^{-1}$ &
   $|\cos\theta| \leq 0.98$
   \\
   EM calorimeter &
   $\sigma_{E}/E = 17\% / \sqrt{E} \oplus 1\%$ &
   $|\cos\theta| \leq 0.99$
   \\
   Hadron calorimeter &
   $\sigma_{E}/E = 45\% / \sqrt{E} \oplus 2\%$ &
   $|\cos\theta| \leq 0.99$
   \\
   \hline
  \end{tabular}
}
\caption{\small The detector parameters used in our simulation study.}
 \label{tb:GLD}
\end{table}

The signal and SM events have been generated by Physsim~\cite{physsim}. The initial-state radiation and bremsstrahlung have been included in the event generations. The beam energy spread was set to be 0.14\% for electron and 0.07\% for positron beams. The finite crossing angle between the electron and positron beams was assumed to be 14 mrad. In the event generations, helicity amplitudes were calculated using the HELAS (HELicity Amplitude Subroutines) library~\cite{helas}, which allows us to deal with the effect of gauge boson polarizations properly. The phase space integration and the generation of parton 4-momenta have been performed by a nubmerical integration program package , BASES/SPRING~\cite{bases}. The parton showering and hadronization have been carried out by using the event generator PYTHIA6.4~\cite{pythia}, where final-state tau leptons are decayed by the MC particle decay software package, TAUOLA~\cite{tauola} to handle their polarizations correctly.

The generated Monte Carlo events have been passed to a detector simulator called JSFQuickSimulator, which implements the geometry and other detector-performance related parameters of an ILC detector concept called GLD~\cite{glddod}. In the detector simulator, hits by charged particles at the vertex detector and track parameters at the central tracker are smeared according to their position resolutions, taking into account correlations due to off-diagonal elements in the error matrix. Since calorimeter signals are simulated in individual segments, a realistic simulation of cluster overlapping is possible. Track-cluster matching is performed for the hit clusters in the calorimeter in order to form pseudo Particle Flow Objects (pPFO), thereby achieving the best attainable jet energy measurements. The resultant detector performance in our simulation study is summarized in Table~\ref{tb:GLD}.

\section{Results from Simulation Study}
\label{sec:analysis}

\begin{figure}
\begin{center}
  \includegraphics[width=10cm]{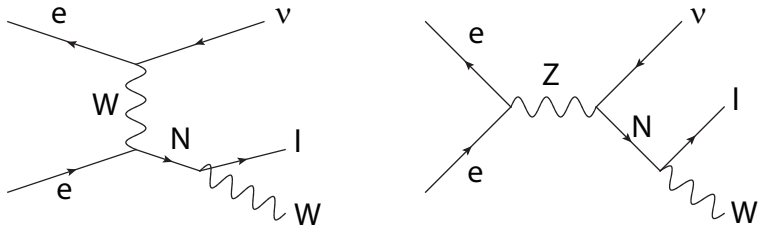}
\end{center}
\caption{\small The diagrams of signal processes, $e^+ e^- \to \nu N$ ($N \to \ell W$).} 
 \label{fig:diagram}
\end{figure}

The simulation has been performed at $\sqrt{s} = 500$ GeV for the first KK mode of right-handed neutrinos, and at $\sqrt{s} = 1$ TeV for the KK modes up to the third level, with an integrated luminosity of 500~fb$^{-1}$ each. We have considered the case with no beam polarization. We have evaluated the measurement accuracies of the masses of KK right-handed neutrinos and their production cross-sections using the processes as shown in Fig.~\ref{fig:diagram} for the cases of the normal, inverted, and degenerate hierarchies of neutrino masses.

\begin{table}
\center{
\begin{tabular}{|l|c|r|r|r|r|r|r|}
\hline
\multicolumn{2}{|c|}{} & 
\multicolumn{3}{|c|}{$\sqrt{s} = 500$ GeV} & 
\multicolumn{3}{|c|}{$\sqrt{s} = 1$ TeV} \\ \hline
KK mode & $\ell$ & (N) & (I) & (D) & (N) & (I) & (D) \\
\hline
1st [fb] & $e$ & 6.524 & 297.5 & 257.1 & 7.79 & 355 & 307 \\
         & $\mu$ & $\cdots$ & $\cdots$ & $\cdots$ & $\cdots$ & $\cdots$ & $\cdots$ \\
         & $\tau$ & 5.490 & 4.176 & 0.113 & $\cdots$ & $\cdots$ & $\cdots$ \\ \hline
2nd [fb] & $e$ & 0.065 & 2.975 & 2.571 & 0.51 & 23.6 & 20.4 \\ \hline
3rd [fb] & $e$ & $\cdots$ & $\cdots$ & $\cdots$ & 0.085 & 3.86 & 3.34 \\ \hline 
\end{tabular}
}
\caption{\small Cross sections of $e^+ e^- \to \nu N$ ($N \to \ell W, W \to q\bar{q}$) at $\sqrt{s} = 500$ GeV and 1 TeV\@. The three dots mean that the cross sections are too low to be explored at the ILC.}
\label{tb:xsec}
\end{table}

\subsection{Study at $\sqrt{s} = 500$ GeV}

Based on the cross sections shown in Table~\ref{tb:xsec}, we have studied $e^+ e^- \to \nu N_1$ ($N_1 \to e W$) for all the neutrino mass hierarchies and $e^+ e^- \to \nu N_1$ ($N_1 \to \tau W$) for the normal and inverted mass hierarchies. In the analysis, we have used the hadronic-decay modes of $W$, which allow us to fully reconstruct the mass of $N_1$.

\subsubsection{Analysis of $\nu N_1 \to \nu e W$}

In the signal event, an isolated-electron track from the decay of $N_1$ is expected. We have therefore selected the electron track and reconstructed two jets in the final state of the signal. Since the isolated-electron track has no energy around it, while a track from a jet has some energy, we have selected tracks with the energy around them within 20 degrees below 5 GeV\@. Then, in the tracks which satisfy the requirement, the track with the maximum energy has been selected as a candidate of the electron track. After picking up the isolated-electron track, the clustering of the jets has been performed. The pPFOs have been combined to form a jet if the two clusters satisfy $y_{ij} < y_{\mathrm{cut}}$, where the variable $y_{ij}$ is defined as
\begin{equation}
y_{ij} = \frac{2 E_i E_j (1 - \cos \theta_{ij})}{E_{\mathrm{vis}}^2}.
\end{equation}
Here, $\theta_{ij}$ is the angle between two clusters, $E_{i (j)}$ are their energies, and $E_{\mathrm{vis}}$ is the total visible energy. All events are forced to have two jets by adjusting $y_{\mathrm{cut}}$. The mass of $N_1$ was, then, reconstructed using the candidate for the electron track and two reconstructed jets. 

\begin{figure}
  \begin{center}
    \includegraphics[width=6cm]{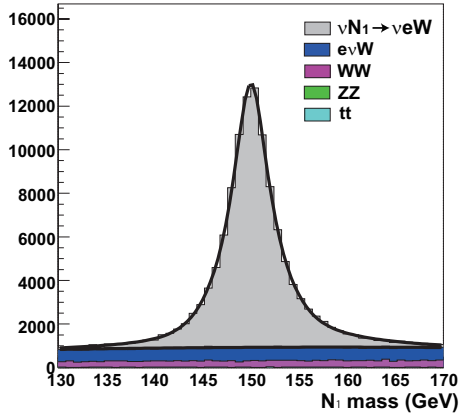}
  \end{center}
  \caption{\small Distribution of the $N_1$ mass reconstructed from $M_{ejj}$ in the case of inverted neutrino mass hierarchy at $\sqrt{s} = 500$ GeV.}
 \label{fig:nmass_emode}
\end{figure}

We have considered the processes shown in Table~\ref{tb:cut_summary} as the background. In order to suppress these background processes, we have applied the following requirement. The isolated electron candidate selected above would have a higher energy then tracks from jets, if it were really coming from the decay of $N_1$. We, hence, required the energy of the isolated-electron candidate ($E_e$) to be 10 GeV $< E_e <$ 200 GeV\@. In addition, since the reconstructed di-jet mass ($M_{jj}$) should be consistent with the $W$ hypothesis for a signal event, we have selected events with 60 GeV $< M_{jj} <$ 100 GeV\@. For reconstruction of the $N_{1}$ mass with the isolated electron and $W$ candidates ($M_{ejj}$), the energy and momentum of the $W$ candidate was corrected to have $W$ mass. Figure \ref{fig:nmass_emode} shows the distribution of $M_{ejj}$ after all the selection cuts in the case of the inverted neutrino mass hierarchy. With the signal region defined by 135 GeV $< M_{ejj} <$ 165 GeV, the numbers of signal and background events before and after selection cuts are summarized in Table~\ref{tb:cut_summary}.

We have evaluated the measurement accuracy of the $N_1$ mass and its production cross section by fitting the $M_{ejj}$ distribution. The $N_1$ mass values and errors from the fit are $150.2 \pm 0.20$ GeV, $150.0 \pm 0.01$ GeV, and $150.0 \pm 0.01$ GeV, for the normal, inverted, and degenerate neutrino mass hierarchies, respectively. These results correspond to the measurement accuracies of 0.14\%, 0.01\%, and 0.01\%. On the other hand, the cross section of $e^+ e^- \to \nu N_1$ ($N_1 \to e W$) events turned out to be measurable with accuracies of 6.5\%, 0.4\%, and 0.4\%. Since the cross-section of the normal hierarchy is much smaller than that of the other hierarchies as shown in Table \ref{tb:xsec}, the measurement accuracy for the normal hierarchy is worse. The results are summarized in Table \ref{tb:resol_summary}.

\subsubsection{Analysis of $\nu N_1 \to \nu \tau W$}

In order to analyze $e^+ e^- \to \nu N_1$ ($N_1 \to \tau W$) events, we have reconstructed all events as 3-jet. A jet with the smallest number of the tracks was assumed to be a tau-jet. The previous signal $e^+ e^- \to \nu N_1$ ($N_1 \to e W$) has been considered as background together with $e^+ e^- \to e \nu W$ and $WW$ shown in Table~\ref{tb:cut_summary}. The following selection cuts have been applied to reduce these backgrounds. The energy of the tau-jet ($E_\tau$) was required to satisfy 10 GeV $< E_\tau <$ 150 GeV in order to reject high energy electron- and muon-tracks from the leptonic-decay modes of $W$ in the $e^{+}e^{-} \to WW$ events. We required the same criteria for the di-jet mass, $M_{jj}$, as in the study of $\nu N \to e W$. Then, the reconstructed $N_1$ mass ($M_{\tau jj}$) was required to be 80 GeV $< M_{\tau jj} <$ 160 GeV.

After applying the selection cuts, a likelihood analysis has been performed. Since the processes $e^+ e^- \to e \nu W$, $\nu N_1$ ($N_1 \to e W$) dominate in the background, we constructed two likelihood functions to separate the signal from $e\nu W$~($\mathcal{L}_{e\nu W}$) and $\nu N_1 \to \nu eq\bar{q}$~($\mathcal{L}_{\nu N_1 \to \nu eq\bar{q}}$). As the input variables of the likelihood functions, we used the number of tracks in the jets of the tau candidate, the energy of the track with the maximum energy in the tau-jet ($E_{\mathrm{max}}$), and the energy of tau-jet with $E_{\mathrm{max}}$ subtracted from it. These likelihood functions were prepared for the normal and inverted neutrino mass hierarchies separately, because we would be able to identify the neutrino mass hierarchies by using the cross section of $\nu N \to \nu e W$ events. We have required $\mathcal{L}_{e\nu W} > 0.79 \ (0.63)$ and $\mathcal{L}_{\nu N_1 \to \nu e q\bar{q} } > 0.13 \ (0.11)$ for the normal (inverted) neutrino mass hierarchy to maximize the signal significance.

\begin{figure}
  \begin{center}
    \includegraphics[width=6cm]{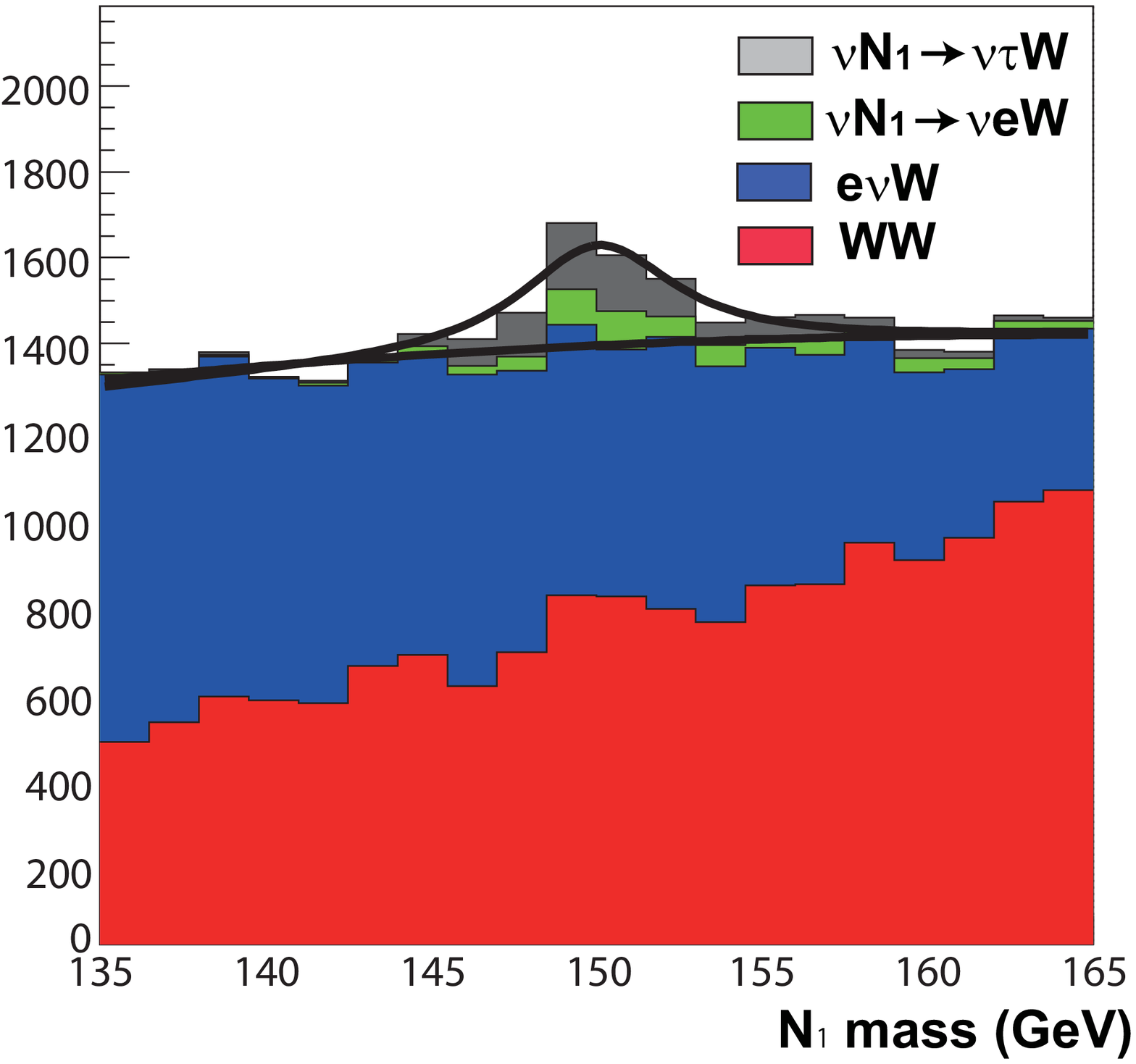}
  \end{center}
  \caption{\small Distribution of the $N_1$ mass reconstructed from $^{\mathrm{col}}M_{\tau jj}$ in the case of inverted neutrino mass hierarchy at $\sqrt{s} = 500$ GeV.}
 \label{fig:nmass_taumode}
\end{figure}

The resolution of the $N_{1}$ mass can be improved by compensating for the missing energy of the $\tau$ decay as follows: Since the $N_{1}$ mass has already been measured through the analysis of $\nu N \to \nu e W$ events, we  can calculate energy of the $N_{1}$ assuming its two-body kinematics. Then, we calculated the energy of the tau as the calculated $N_{1}$ energy minus the energy of the $W$ candidate. We assumed that the direction of the tau coincided the direction of the tau-jet. The corrected mass of $N_{1}$ ($^{\mathrm{col}}M_{\tau jj}$) was reconstructed by using the estimated tau energy and momentum and those of the $W$ candidate. Figure~\ref{fig:nmass_taumode} shows the distribution of $^{\mathrm{col}}M_{\tau jj}$ for the inverted neutrino mass hierarchy after applying all selection cuts. The number of events in the signal region, 135 GeV $< ^{\mathrm{col}}M_{\tau jj} <$ 165 GeV, before and after the selection cuts are summarized in Table~\ref{tb:cut_summary}. Fitting the $^{\mathrm{col}}M_{\tau jj}$ distribution, we have obtained the $N_1$ mass as $149.8 \pm 0.24$ GeV and $150.0 \pm 0.32$ GeV, corresponding to the measurement accuracies of 0.16\% and 0.21\% for the normal and inverted neutrino mass hierarchies, respectively. The cross section of $\nu N_1 \to \nu \tau W$ could be determined with accuracies of 11.3\% and 12.4\% for the two mass hierarchies. These results are summarized in Table~\ref{tb:resol_summary}.

\subsection{Study at $\sqrt{s} = 1$ TeV}

Not only the first but also the second and the third KK modes of right-handed neutrinos can be produced at the ILC with $\sqrt{s} = 1$ TeV\@. Taking into account the cross sections shown in Table~\ref{tb:xsec}, we have studied the process $e^+ e^- \to \nu N_1$ ($N_1 \to \nu e W$) for all the neutrino mass hierarchies and $e^+ e^- \to \nu N_{2,3}$ ($N_{2,3} \to \nu e W$) processes for the inverted and degenerate neutrino mass hierarchies. The background processes considered in this study are shown in Table~\ref{tb:cut_summary_1tev}.

After the selection of the electron track and the reconstruction of two jets with the same procedure as the study at $\sqrt{s} = 500$ GeV, the following selection cuts were applied. To remove electron tracks from jets, we selected high-energy electrons by requiring 10 GeV $< E_e <$ 600 GeV\@. The di-jet mass was required to satisfy the same criteria as the study at $\sqrt{s} = 500$ GeV. Since the $WW$ events have a peak at 500 GeV for the di-jet energy distribution, we have required the di-jet energy should be below 460 GeV\@. Shifting the energy and momentum of the $W$ candidate to have the nominal $W$ mass, the $M_{ejj}$ was calculated. Finally, the signal region has been defined as 135 GeV $< M_{ejj} <$ 165 GeV, 425 GeV $< M_{ejj} <$ 475 GeV, and 720 GeV $< M_{ejj} <$ 780 GeV for the 1st, 2nd, and 3rd KK modes, respectively. The numbers of events before and after the selection cuts are summarized in Table~\ref{tb:cut_summary_1tev}. 

\begin{figure}
 \begin{center}
  \includegraphics[width=12cm]{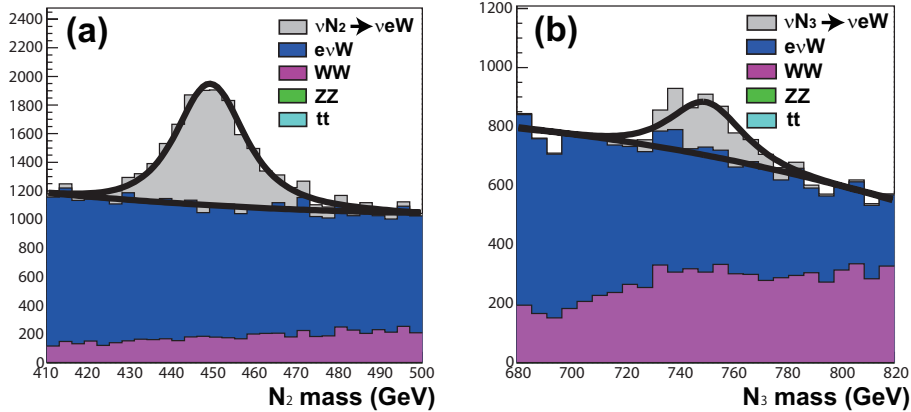}
 \end{center}
 \caption{\small Distributions of the masses of (a) the second and (b) third KK modes reconstructed from $M_{ejj}$ in the case of inverted neutrino mass hierarchy at $\sqrt{s} =$ 1 TeV.}
 \label{fig:nmass_1tev}
\end{figure}

Figure \ref{fig:nmass_1tev} shows the distributions of $M_{ejj}$ for the 2nd and the 3rd KK modes of right-handed neutrinos in the case of inverted neutrino mass hierarchy. Fitting the $M_{ejj}$ distributions, we have obtained the measurement accuracies of $N_{1,2,3}$ masses and their cross sections as shown in Table~\ref{tb:resol_summary}. The masses of right-handed neutrinos could be determined with the accuracy better than 1\%.

\begin{table}
\begin{tabular}{|lc|r|r|rr|}
\hline
Process ($\sqrt{s} =$ 500 GeV) & & Cross sec. & Events No. & \multicolumn{2}{|r|}{Events after cuts} \\
\hline
$\nu N_{1} \to \nu eW$ ($W\to q\bar{q}$) & (N) &   6.5 [fb] &     3,262 &   \multicolumn{2}{|r|}{2,216} \\
                                         & (I) & 297.5 [fb] &   177,700 & \multicolumn{2}{|r|}{101,057} \\
                                         & (D) & 257.1 [fb] &   128,564 & \multicolumn{2}{|r|}{87,588}  \\
\hline 
$e \nu W \to e \nu q\bar{q}$             &     & 4,462 [fb] & 2,231,000 &  \multicolumn{2}{|r|}{22,594} \\
$WW \to \ell \nu q\bar{q}$               &     &   1320 [fb] &   660,000 &   \multicolumn{2}{|r|}{11,324} \\
$ZZ \to \nu \nu q\bar{q},\,
        \ell \ell q\bar{q}$              &     &   108 [fb] &    54,000 & \multicolumn{2}{|r|}{1} \\
$t\bar{t}$                               &     &   531 [fb] &   265,500 & \multicolumn{2}{|r|}{568} \\
\hline
\multicolumn{5}{c}{} \\
\hline
Process ($\sqrt{s} =$ 500 GeV) & & Cross sec. & Events & \multicolumn{2}{|r|}{Events after cuts} \\
\hline
$\nu N_{1} \to \nu\tau W$ ($W\to q\bar{q}$) & (N) & 5.49 [fb] & 2,745 & \multicolumn{2}{|r|}{1,029} \\
                                            & (I) & 4.18 [fb] & 2,090 & \multicolumn{2}{|r|}{821} \\
\hline
$\nu N_{1} \to \nu e W$ ($W \to q\bar{q}$)  &     &  6.52 [fb] &   3,260 & 495 (N) & 554 (I) \\
$e \nu W \to e \nu q\bar{q}$                &     & 4,460 [fb] & 223,100 & 8,989 (N) & 12,276 (I) \\ 
$WW \to \ell \nu q\bar{q}$                  &     & 3,960 [fb] & 1,980,000 & 13,788 (N) & 14,861 (I) \\
\hline
\end{tabular}
\caption{\small Summary of cuts for $\sqrt{s} = 500$ GeV.}
\label{tb:cut_summary}
\end{table}

\begin{table}
\begin{tabular}{|lc|r|r|r|}
\hline
Process ($\sqrt{s} =$ 1 TeV) & & Cross sec. & Events & Events after cuts \\
\hline
$\nu N_1 \to \nu eW$ ($W\to q\bar{q}$) & (N) &  7.79 [fb] &   3,895 &  1,244 \\
                                       & (I) & 355.0 [fb] & 177,700 & 44,021 \\
                                       & (D) & 307.0 [fb] & 153,540 & 29,106 \\
\hline 
$e \nu W \to e \nu q\bar{q}$           &     & 10,320 [fb] & 5,160,000 & 9238 \\ 
$WW \to \ell \nu q\bar{q}$             &     &  560.6 [fb] &   280,300 &  1,234 \\
$ZZ \to \nu \nu q\bar{q},\,
        \ell \ell q\bar{q}$            &     &  42.79 [fb] &    21,393 &      5 \\
$t\bar{t}$                             &     &  29.43 [fb] &    14,715 &      23 \\
\hline
\multicolumn{5}{c}{} \\
\hline
Process ($\sqrt{s} =$ 1 TeV) & & Cross sec. & Events & Events after cuts \\
\hline
$\nu N_2 \to \nu eW$ ($W\to q\bar{q}$) & (I) &  23.6 [fb] &  11,800 &  6,756 \\
                                       & (D) &  20.4 [fb] &  10,200 &  5,820 \\
\hline 
$e \nu W \to e \nu q\bar{q}$           &     & 10,320 [fb] & 5,160,000 & 24,671 \\ 
$WW \to \ell \nu q\bar{q}$             &     &  560.6 [fb] &   140,150 &  4,858 \\
$ZZ \to \nu \nu q\bar{q},\,
        \ell \ell q\bar{q}$            &     &  42.79 [fb] &    21,393 &      0 \\
$t\bar{t}$                             &     &  29.43 [fb] &    14,715 &      0 \\
\hline
\multicolumn{5}{c}{} \\
\hline
Process ($\sqrt{s} =$ 1 TeV) & & Cross sec. & Events & Events after cuts \\
\hline
$\nu N_3 \to \nu eW$ ($W\to q\bar{q}$) & (I) &   3.86 [fb] &     1,932 &  1,131 \\
                                       & (D) &   3.34 [fb] &     1,670 &    961 \\
\hline 
$e \nu W \to e \nu q\bar{q}$           &     & 10,320 [fb] & 5,160,000 & 10,510 \\ 
$WW \to \ell \nu q\bar{q}$             &     &  560.6 [fb] &   280,300 &  6,780 \\
$ZZ \to \nu \nu q\bar{q},\,
        \ell \ell q\bar{q}$            &     &  42.79 [fb] &    21,393 &      0 \\
$t\bar{t}$                             &     &  29.43 [fb] &    14,715 &      0 \\
\hline

\end{tabular}
\caption{\small Summary of cuts for $\sqrt{s} = 1$ TeV.}
\label{tb:cut_summary_1tev}
\end{table}

\begin{table}
\begin{tabular}{|lc|r|r|}
\hline
Process ($\sqrt{s} =$ 500 GeV) & & $N$ mass resolution & Cross-section accuracy \\
\hline
$\nu N_1 \to \nu e W$ & (N) & 0.14 [\%] & 6.5 [\%] \\
                      & (I) & 0.01 [\%] & 0.4 [\%] \\
                      & (D) & 0.01 [\%] & 0.4 [\%] \\
\hline 
$\nu N_1 \to \nu \tau W$ & (N) & 0.16 [\%] &  11.3 [\%] \\
                         & (I) & 0.21 [\%] & 12.4 [\%] \\
\hline
\multicolumn{4}{c}{} \\
\hline
Process ($\sqrt{s} =$ 1 TeV) & & $N$ mass resolution & Cross-section accuracy \\
\hline
$\nu N_1 \to \nu e W$ & (N) & 0.41 [\%] & 13.6 [\%] \\
                      & (I) & 0.01 [\%] &  0.6 [\%] \\
                      & (D) & 0.02 [\%] &  0.7 [\%] \\
\hline 
$\nu N_2 \to \nu e W$ & (I) & 0.05 [\%] & 2.8 [\%] \\
                      & (D) & 0.08 [\%] & 3.1 [\%] \\
\hline
$\nu N_3 \to \nu e W$ & (I) & 0.21 [\%] & 9.9 [\%] \\
                      & (D) & 0.23 [\%] & 10.0 [\%] \\
\hline
\end{tabular}
\caption{\small Summary of measurement accuracies.}
\label{tb:resol_summary}
\end{table}

\section{Discussion}
\label{sec:discussion}

We found that the observation of the higher KK modes, especially the mass spectrum of these particles is measured accurately at the ILC, which allows us to confirm that physics behind these signals is based on higher-dimensional theory. The observation of the masses of KK modes, however, does not directly mean that the signals are coming from physics responsible for neutrino masses and mixings, because there are many scenarios in the framework of higher-dimensional theory which are not related to neutrinos but to other issues such as hierarchy problem and dark matter. 

In order to confirm that the signals are from physics of neutrino masses and mixings through the higher-dimensional theory, we should observe not only the masses of KK modes but also other quantities which depend strongly on parameters of neutrinos. One of such parameters is $\delta_M$, because it is the origin of tiny neutrino masses. However, the smallness of the parameter inevitably leads to the smallness of lepton-number violation in the scenario, so that it is difficult to observe the quantities related to this parameter through processes violating the lepton-number.

\begin{figure}
\begin{center}
  \includegraphics[width=15cm]{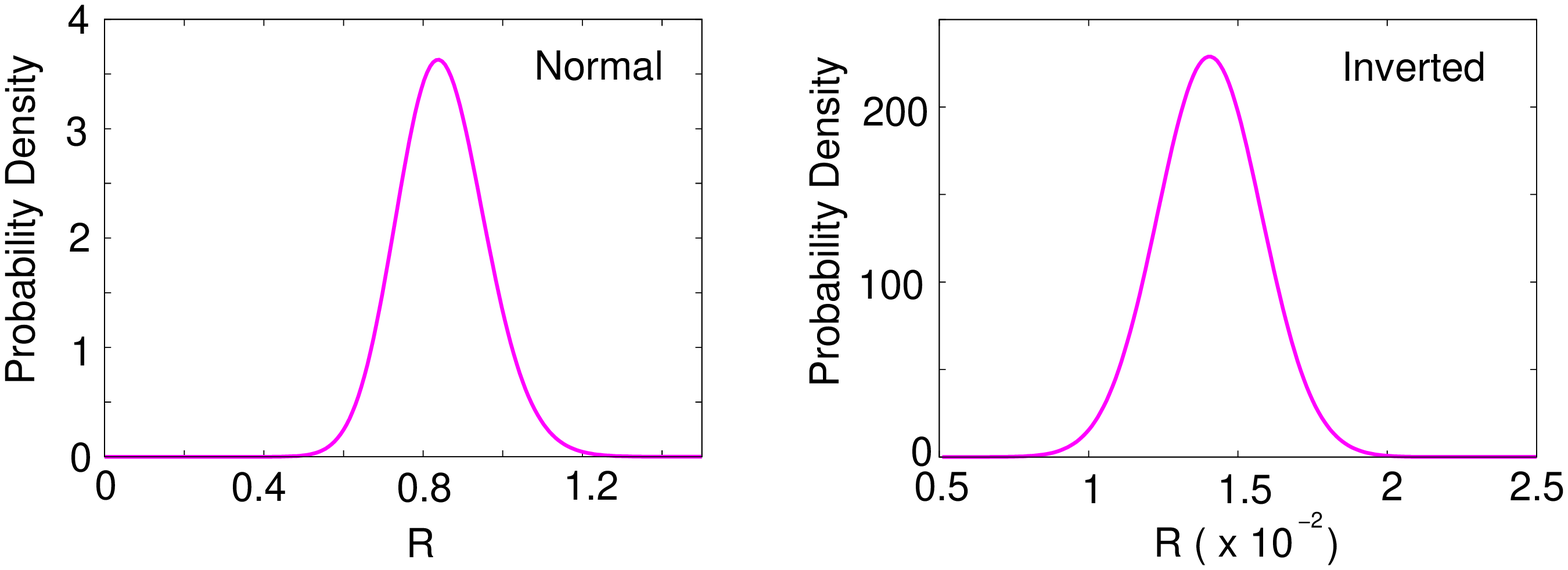}
\end{center}
\caption{\small Probability Densities to determine the branching ratio between the $eW$ and $\tau W$ decay modes of the right-handed neutrino $N_1$. The left figure shows the density for the case of normal mass hierarchy, while right one is for inverted mass hierarchy.} 
 \label{fig:ratio}
\end{figure}

Another interesting parameter is $U_{\rm MNS}$ describing phenomena of neutrino mixings. The higher-dimensional theory for neutrinos considered in this article predicts a specific flavor-structure at the right-handed neutrino sector, and it is completely determined by the mixing parameter $U_{\rm MNS}$ and neutrino masses $M_\nu^d$ as shown in Eq.~(\ref{KKinteraction}). This fact leads to that branching fractions of the right-handed neutrino are governed by $U_{\rm MNS}$ and $M_\nu^d$. We have shown that both the $eW$ and $\tau W$ decay modes of $N_1$ can fortunately be measured at the ILC\@. The ratio of these two branching fractions $R \equiv {\rm Br}(\tau W)/{\rm Br}(e W)$ is theoretically given by the formula,
\begin{eqnarray}
R
=
\frac{
\left[U_{\rm MNS}~M_\nu^d~U^\dagger_{\rm MNS}\right]_{13} ^2 
\left[U_{\rm MNS}~M_\nu^d~U^\dagger_{\rm MNS}\right]_{33}
\left[U_{\rm MNS}~(M_\nu^d)^2~U^\dagger_{\rm MNS}\right]_{11}
}
{
\left[U_{\rm MNS}~M_\nu^d~U^\dagger_{\rm MNS}\right]_{11}^3
\left[U_{\rm MNS}~(M_\nu^d)^2~U^\dagger_{\rm MNS}\right]_{33}
},
\end{eqnarray}
which takes a value of 0.850 for the normal mass hierarchy and 1.58 $\times$ 10$^{-2}$ for the inverted mass hierarchy using the representative points of $U_{\rm MNS}$ in Table~\ref{table:MNS}. With the use of the results in Table~\ref{tb:resol_summary}, we investigate how accurately the ratio can be measured at the ILC\@. The result is shown in Fig.~\ref{fig:ratio} for both cases of normal and inverted mass hierarchies, where probability densities for the determination of the ratio are shown. It can be seen that the ratio will be measured with accuracy of $\pm$ 0.110 and $\pm$ 0.17 $\times$ 10$^{-2}$ for the cases of normal and inverted mass hierarchies, respectively. The ILC has, therefore, a potential not only to discover the signal of higher-dimension theory but also to observe the signal related to neutrino oscillation parameters.

\section{Summary}
\label{sec:summary}

We have investigated ILC signals of the seesaw scenario in a five-dimensional extension of the SM, where right-handed neutrinos live in the bulk and the SM particles stay at a four-dimensional boundary. We focused on the production process of KK right-handed neutrinos, $e^+ e^- \to N \nu$ ($N \to \ell W$, $W \to q\bar{q}$), where the masses of KK right-handed neutrinos $N$ can be fully reconstructed. With realistic Monte-Carlo simulations, we found that the masses of KK neutrinos and their production cross sections can be measured accurately at the ILC as summarized in Table~\ref{tb:resol_summary}. In particular, it was found that the mass and production cross section of the first KK right-handed neutrino can be measured accurately for various hierarchies of neutrino masses with the center of mass energy of 500 GeV\@. In addition, it was shown that masses and production cross sections of the second and third KK neutrinos can be measured with the center of mass energy of 1 TeV.

The ILC also allow us to investigate the flavor structure of the higher-dimensional theory using the measurement of the $N_1$ production cross section followed by its various decay modes. Since branching fractions of $N_1$ decay modes are determined by the masses of neutrinos and their mixing matrix $U_{\rm MNS}$, measuring the ratio between these fractions can be directly compared with the results of neutrino oscillation experiments, which give us an important clue to clarify the mechanism to generate neutrino masses in the framework of the higher dimensional theory. Interestingly, the first or third generation lepton in the final state of the $N_1$ decay can be detected at the ILC, while the LHC will be possible to observe that with the first or second lepton. The ILC will therefore be a complementary machine to the LHC to explore physics of neutrinos in the higher-dimensional theory.

\vspace{1.0cm}
\hspace{0.2cm} {\bf Acknowledgments}
\vspace{0.5cm}

The authors would like to thank all the members of the ILC physics subgroup \cite{Ref:subgroup} for useful discussions. This work is supported by the scientific grants from the ministry of education, science, sports, and culture of Japan (No.~20244028, 20540272, 20740135, 21740174, 22011005, 22244021, and 22244031), and also the grant-in-aid for the global COE program "The next generation of physics, spun from universality and emergence" for K.Y.\ and the Grant-in-Aid for the Global COE Program Weaving Science Web beyond Particle-matter Hierarchy from the Ministry of Education, Culture, Sports, Science and Technology of Japan for M.A..

\end{document}